\documentclass[twocolumn, 5p]{elsarticle}
\journal{Computer Physics Communications}

\RequirePackage{hyperref,xcolor,graphicx,amsmath,amssymb,amsfonts,mathtools,bm}
\RequirePackage{siunitx,xspace,subcaption,tabularx}
\usepackage{url}
\usepackage{array}
\usepackage{enumitem}
\usepackage{booktabs}
\usepackage{temf-finite} 

\DeclareGraphicsExtensions{.pdf,.jpeg,.png}
\DeclareMathOperator{\diag}{diag}

\newcommand{\eh}[1]{{\rm #1}}
\newcommand{\bvec}[1]{\mbox{\boldmath$#1$}}
\newcommand{\dif}[2]{\protect\frac{\protect\partial{}#1}{\protect\partial{}#2}}
\newcommand{\ddif}[2]{\protect\frac{\eh{d}#1}{\eh{d}#2}}
\newcommand{\bnabla}{\/{\nabla}}

\newcommand{\ofx}[1]{\left(#1 \right)} 	
\newcommand{\dd}{\mathrm{d}}	

\newcommand{\figref}[2][]{\ifvmode Figure~\ref{#2}#1\xspace \else Fig.~\ref{#2}#1\xspace \fi}
\newcommand{\tabref}[1]{\ifvmode Table~\ref{#1}\xspace \else Tab.~\ref{#1}\xspace \fi}

\newcommand{\scat}[1]{\bvec{#1}_\mathrm{s}}
\newcommand{\inc}[1]{\bvec{#1}_\mathrm{i}}
\newcommand{\IpA}{\mathbf{R}}
\newcommand{\IpS}{\mathbf{I}_\mathrm{s}}

\newcommand{\edgelength}{l}
\newcommand{\edgelengthRed}{\hat{l}}
\newcommand{\facearea}{A}
\newcommand{\faceareaRed}{\hat{A}}

\newcommand{\pbci}{PBCI\xspace}
\newcommand{\reptil}{REPTIL\xspace}
\newcommand{\fsSC}{FREE-SCF\xspace}
\newcommand{\fsCST}{FREE-CST\xspace}
\newcommand{\wakeFMM}{WAKE-FMM\xspace}
\newcommand{\wakeLW}{WAKE-LW\xspace}
\newcommand{\wakeCST}{WAKE-CST\xspace}
\begin{document}

\begin{frontmatter}
\title{A Scattered-Field Formulation for Coupled Geometric Wakefield and Space Charge Field Simulations in Particle Accelerators}

\author{J.~Christ\corref{fn1}}
    \ead{jonas.christ@tu-darmstadt.de}
    \cortext[fn1]{Corresponding author}
\author{E.~Gjonaj}%
\author{H.~De Gersem}%
\affiliation{organization={Technical University of Darmstadt, Institute for Accelerator Science and Electromagnetic Fields (TEMF)},
            city={Darmstadt},
            country={Germany}}
\date{\today}

\begin{abstract}
We propose a self-consistent simulation model for particle beams in accelerators, which includes the impact of electromagnetic wakefields caused by the geometry of the accelerator chamber. The method is based on a scattered-field formulation for the beam-driven Maxwell's equations. The total electromagnetic field seen by the particles is obtained as the solution of two coupled problems: a purely wakefield problem and a space charge field problem, where for each of these problems, specialized and numerically efficient approaches can be used. To assess the accuracy of the method, we compare simulation results with the analytical solution for a relativistic beam in a uniform accelerator pipe. The numerical efficiency of the method is, furthermore, demonstrated in the beam dynamics study of the multi-cell RF photo-gun installed at the SuperKEK collider facility. We show that electromagnetic wakefields have a non-negligible impact on the quality of the generated beam and, therefore, should be taken into account in the design of high-brilliance electron sources. 
\end{abstract}


\end{frontmatter}

\section{Introduction}
\label{subsec:intro}

The design of particle accelerators requires accurate simulations of the particle-beam dynamics. The latter is determined by the externally applied fields provided by accelerator magnets and RF cavities as well as by the internal space charge fields (SCF) and synchrotron radiation. In addition, the electromagnetic wakefield scattered at the walls of accelerator chamber may have a significant impact on the beam dynamics \cite{Ferrario_2014,Dohlus_2017}. In principle, electromagnetic Particle-in-Cell (EM-PIC) simulations (cf., e.g.\ \cite{Bacci_2003,Candel_2006,Pinto_2014}) can account for all of these effects. However, accelerator applications pose a challenging multi-scale problem. For example in electron accelerators, the size of particle bunches varies in the sub-\si{\mm}-range, whereas accelerator structures are several meters long. The need to resolve the \si{\tera\hertz}-fields excited by the beam on a computational mesh leads to extremely short simulation time steps. Newer approaches, such as pseudo-spectral methods \cite{Vay_2013a,Lehe_2016a} and EM-PIC in the Lorentz boosted frame \cite{Vay_2011}, are able to overcome this difficulty. However, these techniques are specifically designed for plasma applications, where the interaction of the particles with the accelerator geometry is negligible.

For these reason, typical accelerator simulations employ specialized models, which account either for SCF or for transient wakefields only. SCF solvers assume a relativistic bunch of particles, which is slowly accelerated in free space. The rest frame of the bunch may be considered to be inertial and, thus, the inter-particle interaction is described by an electrostatic problem in this frame \cite{Qiang_2006}. Consequently, such simulations do not take into account relativistic retardation and radiation effects. Furthermore, this approach neglects the electromagnetic wakefield scattered by the accelerator walls. 
In so-called wakefield solvers, the particle beam is approximated by a rigid current source moving along the accelerator axis with constant velocity \cite{Bane_1983,Weiland_1992,Gjonaj_2006}. The internal SCF interaction between the particles is neglected. Due to this simplification, the full set of Maxwell's equations in the time domain can be solved numerically while taking into account  electromagnetic scattering due to accelerator geometry \cite{Gjonaj_2023}. However, this approach is not applicable for particle beams in nonuniform motion and/or with varying charge density distribution as is the case, e.g., in an electron RF-gun \cite{Kim_1988}. 

In this paper, we describe a methodology for the self-consistent simulation of the electromagnetic field and the beam dynamics problems in particle accelerators. The method is based on the scattered-field formulation of Maxwell's equations for the beam-driven problem \cite{Dehler_1995,Nisiyama_2000,Christ_2024a}. The total electromagnetic field of the particles is obtained by a coupled solution of a wakefield and a SCF problem. This completely avoids the need for interpolation of particle currents on the mesh, which is the computationally most expensive step in EM-PIC simulations \cite{Godfrey_2013,Pinto_2014,Efimenko2015}. Extending our previous work in \cite{Christ_2024a}, we include relativistic radiation and retardation effects in the formulation by employing a Liénard-Wiechert solution for the particle's SCF. Furthermore, the method is extended to handle arbitrarily curved geometric boundaries. We demonstrate the validity and ccuracy of this approach in Section \ref{sec:verification} by a detailed comparison of simulations with the analytical solution for a relativistic electron bunch in a homogeneous beam pipe. Finally, in Section \ref{sec:gun_simulation}, beam dynamics simulations are carried out for a high-current, multi-cell electron photo-gun, which is installed in the SuperKEKB collider \cite{Natsui_2013}, for which we present a comparison with EM-PIC as well.


\section{Beam-driven electromagnetic fields}
\label{sec:theory}

The problem is described by Maxwell's equations in the time domain,
\begin{equation}
\label{maxwell}
\begin{split}
\bnabla\cdot\bvec{D} &=\rho ,\quad \dif{\bvec{D}}{t}-\bnabla\times\bvec{H} =-\bvec{J},
\\
\bnabla\cdot\bvec{B} &= 0,\quad \dif{\bvec{B}}{t}+\bnabla\times\bvec{E} = 0,
\end{split}
\end{equation}
where $\bvec{E}$ and $\bvec{H}$ are the electric and magnetic field strengths with corresponding flux densities, $\bvec{D}=\varepsilon_0 \bvec{E}$ and $\bvec{B}=\mu_0\bvec{H}$, respectively. We consider only the case of electromagnetic fields in vacuum, which is a reasonable assumption for most accelerator applications. Equations \eqref{maxwell} are subject to the boundary conditions,
\begin{equation}
\label{maxwell_bc}
\bvec{n}\times\bvec{E}|_{\Gamma} = 0 ,\quad \bvec{n}\cdot\bvec{B}|_{\Gamma} = 0 ,
\end{equation}
where we assume all accelerator walls $\Gamma$, to be perfectly electrically conducting (PEC).
Finally, for a bunch of $N$ particles of charge $q_p$, positions $\bvec{r}_p(t)$ and velocities $\bvec{v}_p(t)$, the charge and current densities in \eqref{maxwell}  are given respectively by
\begin{align}
\label{maxell_current}
\rho(\bvec{r}, t) = \sum_{p=1}^{N} q_p\delta(\bvec{r}-\bvec{r}_p), \quad
\bvec{J}(\bvec{r}, t) & = \sum_{p=1}^{N} q_p\bvec{v}_p\delta(\bvec{r}-\bvec{r}_p) . \nonumber
\end{align}

To separate the contribution of the particle current from that of the boundary conditions, we make the ansatz
\begin{equation}
\label{maxwell_ansatz}
\bvec{E} = \inc{E} + \scat{E} , \quad \bvec{H} = \inc{H} + \scat{H} ,
\end{equation}
and similarly for the fluxes $\bvec{D}$ and $\bvec{B}$. In \eqref{maxwell_ansatz}, the {\em incident fields} $(\inc{E},\inc{H})$ denote solutions of \eqref{maxwell} with suitable, but arbitrary boundary conditions (for example, in free space). The {\em scattered fields} $(\scat{E},\scat{H})$ account for the contribution of the boundary conditions on $\Gamma$. They are solutions of the homogeneous Maxwell's equations
\begin{equation}
\label{maxwell_scattered}
\begin{split}
\bnabla\cdot\scat{D} &= 0,\quad \dif{\scat{D}}{t}-\bnabla\times\scat{H} = \bvec{0},
\\
\bnabla\cdot\scat{B} &= 0,\quad \dif{\scat{B}}{t}+\bnabla\times\scat{E} = \bvec{0},
\end{split}
\end{equation}
with modified boundary conditions,
\begin{equation}
\label{maxwell_scattered_bc}
\bvec{n}\times\scat{E}|_{\Gamma} = -\bvec{n}\times\inc{E}|_{\Gamma} ,\quad  \bvec{n}\cdot\scat{B}|_{\Gamma} = -\bvec{n}\cdot\inc{B}|_{\Gamma} .
\end{equation}

Equations \eqref{maxwell_ansatz} to \eqref{maxwell_scattered_bc} represent the scattered-field formulation, which provides the basis for the simulation approach used in this paper. The motivation for this is as follows. The charge distribution of a high energy particle bunch in the rest frame of the bunch varies slowly over long propagation distances in the accelerator. Under these conditions, $\inc{E}$ and $\inc{H}$, can be determined, either analytically or by a suitable numerical approximation of Maxwell's equations.
The scattered-field equations \eqref{maxwell_scattered}, on the other hand, do not depend explicitly on the individual particle positions and velocities. Unlike conventional EM-PIC, the numerical solution of these equations does not require interpolating individual particle currents on a mesh. This substantially reduces numerical complexity.

In accelerator physics literature, the electromagnetic fields due to the presence of accelerator walls are usually referred to as wakefields (cf.~\cite{Palumbo_1995}). Since $(\inc{E},\inc{H})$ are arbitrary, this description is not fully identical with our definition of scattered fields in \eqref{maxwell_ansatz}. Nevertheless, in the following, we will use the two notations interchangeably, while always referring to \eqref{maxwell_ansatz}. Similarly, we will refer to $(\inc{E},\inc{H})$ either as incident or as space charge fields. While the latter usually suggests an electrostatic or quasistatic field context, this does not need to be the case in our formulation.

\section{Solution of the wakefield problem}
\label{sec:wakefields}

\subsection{Discretization of Maxwell's equations by FIT}

For the discretization of \eqref{maxwell}, we apply the Finite-Integration Technique (FIT) using a staggered Cartesian mesh \cite{Weiland_1977}. Below, we shortly summarize the method for the reader's convenience as well as to introduce notations. The degrees of freedom of FIT are electric and magnetic voltages, defined as
\begin{align}
\label{fit_dofs}
    \left[\efit\right]_j = \int_{l_j} \bvec{E} \cdot \dd \bvec{l}_j , \quad 
    \left[\hfit\right]_j = \int_{\tilde{l}_j} \bvec{H} \cdot \dd \bvec{\tilde{l}}_j ,
\end{align}
where $\bvec{l}_j$ and $\bvec{\tilde{l}}_j$ denote oriented edges in the primal and dual meshes of FIT, respectively. In addition, discrete magnetic and electric fluxes, $\bfit$ and $\dfit$, are defined as electromagnetic field integrals over the areas of elementary faces on the mesh. With these definitions, the semi-discrete FIT equations read \cite{Weiland_1977, Clemens_2001a},
\begin{equation} \label{eq_FIT_HF}
    \ddif{}{t} \begin{pmatrix} \dfit \\ \bfit \end{pmatrix}
        =  \begin{pmatrix} 0 & \curlfit^T \\ -\curlfit & 0 \end{pmatrix} \begin{pmatrix} \efit \\ \hfit \end{pmatrix} - \begin{pmatrix} \jfit \\ 0 \end{pmatrix} ,
\end{equation}
where $\jfit$ is the discrete current vector defined on the mesh and $\curlfit$ is the {\em curl}-operator of FIT.
Furthermore, the relation between fluxes and voltages is provided by a pair of diagonal {\em material matrices}, $\Meps$ and $\Mmu$, as
\begin{align}
\label{fit_material_matrices}
    \dfit = \Meps \hfit , \quad 
    \bfit = \Mmu \hfit .
\end{align}
These relations embody the approximation of electromagnetic materials and that of the boundary conditions on the mesh \cite{Clemens_2001a}. For example, at a PEC boundary, \eqref{maxwell_bc} is imposed by simply setting entries in $\Meps$ and $\Mmu$ to zero if the corresponding geometric entity of the mesh (a mesh edge or face) is part of the PEC domain.

Considering now the wakefield problem for highly relativistic particles, $\jfit$ moves with a constant speed, $v\sim c$, along the accelerator axis. Thus, the solution of \eqref{eq_FIT_HF} can be obtained within a short computational window, which is co-moving with the particle bunch \cite{Bane_1983}.
This allows to reduce the computational effort required in the simulation of very large accelerator structures. In addition, for the time discretization of \eqref{eq_FIT_HF}, a split-operator technique can be applied, which completely eliminates numerical dispersion in the propagation direction of the bunch (cf.\ \cite{Lau_2005,Gjonaj_2006}). These specialized techniques to the wakefield problem have been implemented a.o.\ in the \pbci-code \cite{Gjonaj_2006} on which the simulations presented in this work are based. For a detailed description of this code, we refer to \cite{Niedermayer_2016,Gjonaj_2023}. 

\subsection{Discrete scattered-field formulation}
\label{subsec:sff-fit}

Applying a decomposition of the FIT degrees of freedom into incident and scattered parts as in \eqref{maxwell_ansatz} the semi-discrete formulation corresponding to \eqref{maxwell_scattered} is obtained as,
\begin{equation}
\label{eq_ScatField_HF}
    \ddif{}{t} \begin{pmatrix} \dfit_{\mathrm{s}} \\ \bfit_{\mathrm{s}} \end{pmatrix}
        = \begin{pmatrix} 0 & \curlfit^T \\ -\curlfit & 0 \end{pmatrix}
        \begin{pmatrix} \efit_{\mathrm{s}} \\ \hfit_{\mathrm{s}} \end{pmatrix} 
        - \begin{pmatrix} 0 \\ \jfit_\mathrm{mag} \end{pmatrix} .
\end{equation}
In \eqref{eq_ScatField_HF}, the magnetic current term, $\jfit_\mathrm{mag}$, accounts for the inhomogeneous boundary condition \eqref{maxwell_scattered_bc}.
Given the known incident field voltages on the mesh, $\efit_\mathrm{i}$, the boundary condition \eqref{maxwell_scattered_bc} dictates $[\efit_\mathrm{s}]_j = - [\efit_\mathrm{i}]_j$ for all edges $j$, which constitute the boundary $\Gamma$ of a PEC domain in the mesh. In the case of Cartesian meshes and a staircase approximation of the geometry, this leads to an equivalent magnetic current given by \cite{Christ_2024a},
\begin{equation}
\label{magnetic_current1}
    \jfit_\mathrm{mag} = \curlfit \IpS \efit_\mathrm{i} ,
\end{equation}
where $\IpS$ is a diagonal interpolation matrix with $[\IpS]_{jj} = -1$ if edge $j$ is a boundary edge and $0$ otherwise. Note that the boundary condition for $\scat{B}$ is automatically fulfilled by virtue of Faraday's law.

All matrix operators in \eqref{eq_ScatField_HF}, \eqref{magnetic_current1} are identical with those of the original FIT formulation for the total field \eqref{eq_FIT_HF}. Thus, the same time-marching technique as well as the moving window approach used in standard wakefield simulations can be applied for the solution of \eqref{eq_ScatField_HF} as well.

\subsection{Boundary conformal approximation}
\label{subsec:scat-iq}

Since scattered fields are excited on the PEC boundary $\Gamma$ of the accelerator structure, the approximation of its geometry is crucial for numerical accuracy. The accuracy of the staircase boundary approximation for the magnetic current \eqref{magnetic_current1} is often insufficient for practical simulations \cite{staircase_2014}. Following the ideas in \cite{Dey_1997,Krietenstein_1998}, below, we introduce a boundary conformal approximation for the discrete magnetic current $\jfit_\mathrm{mag}$ on the mesh.

\begin{figure}[hbt]
    \centering
    \subcaptionbox{}[0.45\linewidth]{
        \includegraphics[scale=0.9]{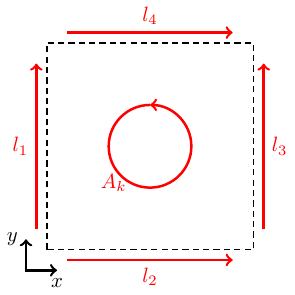}
   }
    \subcaptionbox{}[0.45\linewidth]{
        \includegraphics[scale=0.9]{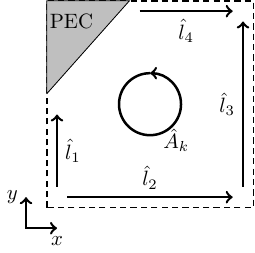}
   }
    \caption{Primal FIT face in free space (a) and partially filled with a PEC material (b). The edge lengths and face areas used in the definition of the FIT degrees of freedom for the boundary conformal approximation are shown.}
        \label{fig_primalFITFaraday_FITmot}
\end{figure}

The method is illustrated in \figref{fig_primalFITFaraday_FITmot} depicting a primary mesh face, partially contained within a PEC region. The scattered electric voltages $[\efit_{\mathrm{s}}]_j$ are defined on the truncated edge paths denoted by $\edgelengthRed_j$, $j=1,\cdots4$, where $\edgelengthRed_j$ corresponds to the portion of the mesh edge $j$ that is contained in the vacuum region. Correspondingly, the scattered magnetic flux $[\bfit_\mathrm{s}]_k$ is defined on the vacuum portion $\faceareaRed_k$ of the area, (see \figref[b]{fig_primalFITFaraday_FITmot}). The incident electric voltages $[\efit_\mathrm{i}]_j$ and magnetic flux $[\bfit_\mathrm{i}]_k$ are defined with respect to the full-edge lengths $\edgelength_j$ and the full-face area $\facearea_k$ respectively. These quantities are depicted in red in \figref[a]{fig_primalFITFaraday_FITmot}.

Integrating \eqref{maxwell_ansatz} over edge $j$ and assuming linear variation of $\inc{E}$ along the edge, it follows for the total voltage,
\begin{equation}
    \left[\efit\right]_j = \left[\efit_\mathrm{s}\right]_j + \frac{\edgelengthRed_j}{\edgelength_j} \left[\efit_\mathrm{i}\right]_j
    .
\end{equation}
Analogously, the total magnetic flux on the face is,
\begin{equation}
    \left[\bfit\right]_k = \left[\bfit_\mathrm{s}\right]_k + \frac{\faceareaRed_k}{\facearea_k} \left[\bfit_\mathrm{i}\right]_k
    .
\end{equation}
These definitions can be extended to the whole domain by introducing two interpolation matrices $\IpA_{\edgelength}=\diag(\edgelengthRed_j / \edgelength_j)$ and $\IpA_{\facearea}=\diag(\faceareaRed_k / \facearea_k)$. Then, Faraday's law for the scattered field reads 
\begin{equation}
    \ddif{}{t} \bfit_{\mathrm{s}} = -\curlfit \efit_{\mathrm{s}} - \curlfit \IpA_\edgelength \efit_\mathrm{i} - \IpA_\facearea \ddif{}{t}\bfit_\mathrm{i} ,
\end{equation}
since the incident field part fulfills \eqref{eq_FIT_HF}, $\ddif{}{t} \bfit_\mathrm{i} = - \curlfit \efit_\mathrm{i}$. Hence, the new scattered-field equations have the same form as in \eqref{eq_ScatField_HF} with a modified magnetic current,
\begin{equation}
\label{eq_ScatIpa_jmag}
    \jfit_\mathrm{mag} = \ofx{\curlfit \IpA_\edgelength - \IpA_\facearea \curlfit}\efit_\mathrm{i} .
\end{equation}

Note that the staircase approximation \eqref{magnetic_current1} is obtained as a special case of \eqref{eq_ScatIpa_jmag} when the mesh fits exactly to the boundary $\Gamma$ as can be the case for a planar geometry. In this case, $[\IpA_{\edgelength}]_{jj}=0$ if edge $j$ is in a PEC region and $[\IpA_{\edgelength}]_{jj}=1$ for a vacuum edge. The same is true for $[\IpA_{\facearea}]_{kk}$ with respect to face $k$.
Thus, $[\curlfit \IpA_\edgelength - \IpA_\facearea \curlfit]_{kj} = - [\curlfit]_{kj}$, only if face $k$ is in vacuum and edge $j$ is in PEC, and (as in \eqref{magnetic_current1}) is $0$ otherwise.

\section{Solution of the space charge field problem}
\label{sec:space-charge}

The computation of $(\inc{E},\inc{H})$ in \eqref{maxwell_ansatz} depends on whether the field evaluation is needed at the particle positions or at the distant walls of the accelerator chamber. The former accounts for direct particle-particle interactions. The latter provides the magnetic current excitation for the scattered-field formulation (\ref{eq_ScatField_HF}). In most cases, it is convenient to choose the SCF of the beam in free space as incident field. Below, we describe the numerical procedures, which we use for the solution of the free-space SCF problem.

\subsection{Particle field computation}
\label{subsec:green}

Most particle tracking codes for accelerator applications such as Astra \cite{Floettmann_2017}, Impact-T \cite{Qiang_2006} and \reptil \cite{Gjonaj_2022} are based on a quasistatic approximation for the SCF \cite{chao-book}. This approximation assumes that the average rest frame of the bunch is Lorentz-inertial, while the energy spread of the particles within the bunch is small. In the case of a particle bunch moving in the $z$-direction with mean Lorentz factor $\gamma_0$, the particle positions and velocities transform in the rest frame as $\bvec{r}_p' = (x_p, y_p, \gamma_0 z_p)$ and $\bvec{v}_p' = (v_{xp}, v_{yp}, 0)$, with $p=1,\dots, N$, where the unprimed coordinates are in the laboratory frame. Thus, Maxwell's equations reduce to an electrostatic field problem in the rest frame of the bunch. The electromagnetic field of the particles in the laboratory frame is obtained by back-transforming the rest-frame fields as
\begin{align}
\label{eq_Lorentz_transform}
E_x &= \gamma_0 E_x' ,  & B_x &= -\sqrt{\gamma_0^2-1} E_y' , \nonumber \\
E_y &= \gamma_0 E_y' ,  & B_y &= \sqrt{\gamma_0^2-1} E_x' , \\
E_z &= E_z',          & B_z &= 0 .\nonumber
\end{align}

For the computation of SCF at the positions of the particles, we employ the  Green's function approach implemented in \reptil. Hereby, the charge density of the particles is interpolated on a mesh enclosing only a small volume around the bunch.
The electrostatic SCF in the rest frame of the bunch is obtained by convolution of the charge density with the free-space Green's function using a 3D-DFT approach on the mesh \cite{Hockney_1965}. The free-space boundary condition is imposed by padding the discrete charge density and Green's function using Hockney's domain doubling technique \cite{Hockney_1988}.

\subsection{Boundary field computation}
\label{subsec:fmm}

To obtain the magnetic current for the scattered-field formulation \eqref{eq_ScatField_HF}, the SCF of the particles is needed also at the boundary $\Gamma$ of the accelerator chamber. For this, we may adopt again a quasistatic field approximation reducing to a Poisson problem in the rest frame of the bunch. However, due to the large distance between particles and accelerator walls, the 3D-DFT approach requires a huge computational mesh, which is not feasible for practical simulations. Instead, we employ a mesh-free approach based on the Fast Multipole Method (FMM) \cite{Greengard_1988}. The method uses an adaptive tree algorithm to compute a suitable multipole decomposition of the Coulomb field in the rest frame of the bunch. Using FMM, the SCF of an $N$-particle bunch can be evaluated at arbitrary positions with numerical complexity $\mathcal{O}(N)$ \cite{Greengard_1988}. For our specific purpose of incident field computation at the boundary $\Gamma$, the method is particularly efficient. Since the distance of accelerator walls from the bunch is typically much larger than the inter-particle distance, only a few multipole terms in the FMM expansion are needed for sufficient accuracy in the boundary field evaluation.

The application of FMM in \reptil is based on the RecFMM library \cite{Zhang_2016}. It implements an asymmetric version of FMM, which allows for the construction of two independent, source and target trees, respectively. This is crucial for our specific purpose of boundary field computation. In the \reptil implementation, the source tree corresponds to the bunch particles, whereas the target tree includes the end points of boundary edges in the FIT mesh, from which the incident field voltages  $\efit_\mathrm{i}$ on $\Gamma$ are obtained from \eqref{fit_dofs} using the trapezoidal rule. The numerical efficiency of this approach is enhanced by the application of exponential expansions for the multipole-to-local transformations \cite{Greengard_1997} as well as by the recursive parallelization scheme employed in RecFMM \cite{Zhang_2016}. 

\subsection{Fully relativistic case}
\label{subsec:lw}

The quasistatic approximation for the incident field of the particles neglects relativistic radiation and retardation. These effects become important in situations such as, e.g., electron acceleration in an RF photo-gun \cite{relativistic-gun,retardation_2009}, where the local rest frame of the bunch is far from inertial. In order to account for this, we employ a Liénard-Wiechert (LW) approach for the computation of the incident boundary fields. The fields of a relativistic particle moving along a path, $\bvec{r}_p=\bvec{r}_p(t)$, at an observation point $\bvec{r}$ are \cite{Jackson_1999},
\begin{equation}
\label{lw}
\begin{split}
&
\bvec{E}(\bvec{r},t) =
\frac{
q\frac{\bvec{n}_p-\bvec{\beta}_p}{\gamma_p^2 |\bvec{r}-\bvec{r}_p|^2}+
q\frac{\bvec{n}_p\times\left[(\bvec{n}_p-\bvec{\beta}_p)\times\dot{\bvec{\beta}}_p\right]}{c|\bvec{r}-\bvec{r}_p|}    
}
{4\pi\varepsilon_0 (1-\bvec{n}_p\cdot\bvec{\beta}_p)^3}
\\
&
\bvec{B}(\bvec{r},t) = \frac{1}{c}\bvec{n}_p\times\bvec{E}(\bvec{r},t) ,
\end{split}
\end{equation}
where $q$ is the charge of the particle, $\bvec{\beta}_p=\dot{\bvec{r}}_p/c$, $\gamma_p = (1-\bvec{\beta}_p^2)^{-1/2}$, and $\bvec{n}_p$ is the unit vector pointing from the particle to the observer. All $p$-quantities are evaluated at the retarded time $t_r$, which is defined by
\begin{equation}
\label{lw_retarded}
c(t-t_r) = |\bvec{r}-\bvec{r}_p(t_r)| .
\end{equation}
It is clear from \eqref{lw} and \eqref{lw_retarded} that the evaluation of LW-fields requires the full $N$-particle trajectories to be stored in memory. The number of operations needed to obtain the incident voltages $\efit_\mathrm{i}$ scales with the number of source particles times the number of boundary points on the FIT mesh. Hence, this approach is computationally much more demanding than the quasistatic FMM-approach described in Section \ref{subsec:fmm}. 

\section{Putting the pieces together}
\label{sec:coupling}

\begin{figure}[htb]
    \centering
    \includegraphics[width=0.98\linewidth]{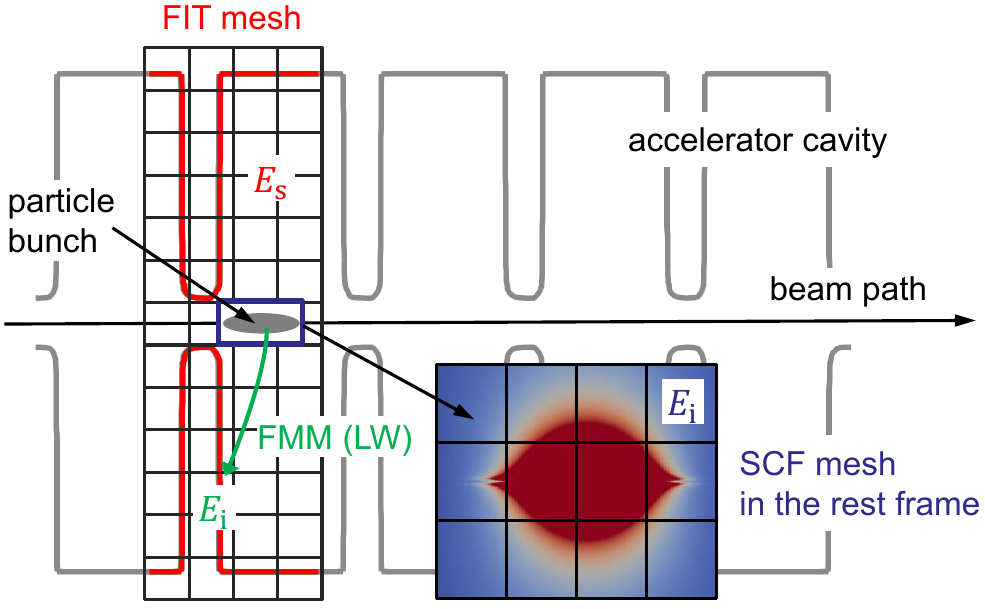}
    \caption{Schematic view of the coupling procedure depicting a particle bunch, an accelerator structure and the two meshes used for the SCF and wakefield computations, respectively.
    } \label{fig_coupling}
\end{figure}

The field coupling procedure to obtain the total electromagnetic field \eqref{maxwell_ansatz} at the particle positions is shown schematically in \figref{fig_coupling}. The procedure uses two different computational meshes. In every time step of the simulation, the SCF $(\inc{E}, \inc{H})$, in the near-particle range is computed on the small mesh (shown in blue) tightly enclosing the particle bunch. This allows to resolve charge density variations within the bunch, which are responsible for the direct space charge interaction between the particles. For the field computation, the quasistatic field approximation based on the 3D-DFT approach with free-space boundary conditions described in Section \ref{subsec:green} is used.

A second mesh is used to solve the FIT-equations \eqref{eq_ScatField_HF} for the scattered fields shown in red in the Figure. This mesh moves along the beam path with the speed of light $c$. This allows for dispersion-free integration of \eqref{eq_ScatField_HF} using the operator-splitting method for wakefield simulations implemented in \pbci \cite{Gjonaj_2006}. Since wakefields are collectively excited by the bunch, this mesh does not need to resolve the individual particle positions. Therefore, the maximum stable time step can be substantially larger than in a conventional EM-PIC simulation. However, the mesh needs to be large enough in order to include the relevant accelerator geometry in the discretization.

The coupling between the two solvers is realized by the magnetic boundary current $\jfit_\mathrm{mag}$ in \eqref{eq_ScatField_HF}. The incident fields $(\inc{E}, \inc{H})$ on $\Gamma$ are computed either with the FMM or by the LW approach \eqref{lw}, \eqref{lw_retarded} as discussed in Section \ref{subsec:fmm} and \ref{subsec:lw}, respectively. Hereby, the fields are evaluated only on those edges of the FIT mesh, which define the geometric boundary of the accelerator structure. Given the incident voltages $\efit_\mathrm{i}$, the current $\jfit_\mathrm{mag}$ is determined by the boundary conformal approximation \eqref{eq_ScatIpa_jmag}. This procedure is depicted in green in \figref{fig_coupling}. 

Finally, the total particle fields are obtained by interpolating the SCF as well as the wakefield solution at the particle positions in the respective meshes. This procedure is repeated in every time step of the simulation, where the particle positions and momenta are updated according to the resulting Lorentz force acting on them. We refrain from discussing the time stepping procedure as this goes beyond the scope of this paper. Note, however, that the field-coupling procedure does not enforce any restriction on the settings of the individual solvers involved in the computation. This allows to reuse the optimization strategies developed for beam dynamics as well as for wakefield simulations, which are implemented in specialized accelerator codes such as \reptil and \pbci. More importantly, there is no need to interpolate the current of individual particles on the mesh. This procedure is by far the computationally most expensive step in EM-PIC, since a strictly charge-conserving current interpolation scheme is required for the numerical stability of these simulations (cf., e.g., the Boris method \cite{Boris_1970} and see also \cite{Godfrey_2013,Pinto_2014}).

\section{Numerical verification and accuracy}
\label{sec:verification}

\subsection{Particle beam in a uniform pipe}
\label{subsec:sc-impedance}

We consider a rigid particle bunch with constant energy moving along the axis of an infinitely long pipe of uniform cross-section. The electromagnetic wakefield is due to the mirror charge of the bunch at the PEC walls of the pipe. Since the geometry of the pipe is otherwise uniform, this is (somewhat misleadingly) referred to in the literature as a space charge wakefield. For a pencil-like bunch of total charge $Q$ and Lorentz-factor $\gamma$, the longitudinal electric field component on-axis is \cite{Chao_1993,Ng_1984}
\begin{align}
\label{eq_validation_EzTotal}
    E_{z}\ofx{s} = 
    -\frac{Q}{2 \pi \varepsilon_0 \gamma^2} \Lambda \ddif{\lambda}{s} ,
\end{align}
where $s=z-z_0$ denotes the distance of the observation point from the bunch center, $z_0=z_0(t)$, $\lambda(s)$ is the line charge density and $\Lambda$ a geometric form factor. For circular and rectangular beam pipes, the latter is given by
\begin{equation}
\label{eq_validation_Lambda}
    \Lambda = \begin{cases}
    \ln \frac{b}{a} + \frac12, & \text{circular} \\
    \ln \left(\frac{2h}{\pi a} \tanh \left(\frac{\pi w}{2 h} \right) \right) + \frac12, & \text{rectangular} , \\
    \end{cases}
\end{equation}
where $a$ is the beam radius and $b$ the radius of the circular pipe. Furthermore, $h$ and $w$ (with $w \gg h$) are, respectively, the height and width of the pipe cross-section in the rectangular case \cite{Ng_1984}.

The closed form solution \eqref{eq_validation_EzTotal} represents the steady-state wakefield, which settles in the pipe for long bunch propagation distances. Furthermore, it assumes a long bunch in the beam rest frame, i.e.,\ $\sigma_z \gg b / \gamma$ (or $\sigma_z \gg h / \gamma$ in the rectangular case). Nevertheless, this solution provides a unique possibility to verify our numerical simulations. Furthermore, from \eqref{eq_validation_EzTotal} we can separate the free-space SCF contribution. This is given by \cite[eq.\ (11.22)]{Stupakov_2018}
\begin{align}
\label{eq_validation_EzFreeSpace}
    E_{\mathrm{i},z}\ofx{s} &=-\frac{Q}{2 \pi \varepsilon_0 a^2} 
    \int_{-\infty}^{\infty} \dd s^{\prime} \lambda\ofx{s^{\prime}} \nonumber \\ 
    & \left(\frac{s-s^\prime}{\sqrt{a^2 / \gamma^2+\left(s-s^{\prime}\right)^2}}-\frac{s-s^\prime}{\left|s-s^{\prime}\right|}\right)  ,
\end{align}
where the expression can be evaluated numerically for arbitrary $\lambda(s)$. Thus, the purely geometric wakefield is $E_{\mathrm{s},z} = E_{z} - E_{\mathrm{i},z}$. This quantity can be directly compared with the simulation results obtained by the numerical solution of \eqref{eq_ScatField_HF}. Note that in all these derivations, a rigid particle bunch is assumed. Thus, the verification procedure does not include the back-coupling of the wakefields to the beam dynamics.

\subsection{Model setup}
\label{subsec:validationModel}

We consider a Gaussian electron bunch of rms length $\sigma_z = \SI{10}{\mm}$, radius $a=\SI{0.5}{\mm}$ and charge $Q = \SI{1}{\nano\coulomb}$. The particles move strictly along the $z$-direction with a kinetic energy of $E_{\mathrm{kin}}=\SI{15}{\MeV}$ corresponding to a Lorentz factor $\gamma \approx 30$. At this energy, electrons are highly relativistic with $v_{pz} \approx 0.9995 c$. In order to match the assumptions for \eqref{eq_validation_EzTotal}, we inject the particles at $z=0$ into a semi-infinite pipe and let them propagate until a steady-state electromagnetic field settles in the moving computational window. In the simulations, we consider both, cylindrical and rectangular beam pipes of different sizes.

Due to the way the bunch is injected, transient waveguide modes are initially excited, which do not contribute to the steady-state solution \eqref{eq_validation_EzTotal}. These modes are either evanescent or propagate in the pipe with group velocity $v_\mathrm{G}<c$. Thus, their contribution within the computational window decays with increasing propagation distance. The necessary distance $l_\mathrm{t}$ that the bunch has to travel until the steady-state is obtained can be estimated from problem parameters.
Given the spectral width of the bunch, $\sigma_\omega = c / \sigma_z$, we expect waveguide modes excited in the frequency range $\omega \lesssim n \sigma_\omega$, with $n = 10$ as a reasonable choice. The fastest mode with non-zero longitudinal electric field is the TM ground mode with group velocity $v_\mathrm{G} = c \sqrt{1-(\omega_c / n \sigma_\omega)^2}$, where $\omega_c$ is the cutoff frequency of the mode. Denoting the length of the computational window by $l_\mathrm{w}$, the bunch has to travel a distance of at least $l_\mathrm{t} = c / (c - v_\mathrm{G}) l_\mathrm{w}$ to fully overtake this mode, when excited at the front of the moving window. 

\subsection{Steady-state solution}

\begin{figure}[htb]
    \centering
    \includegraphics[width=0.85\linewidth]{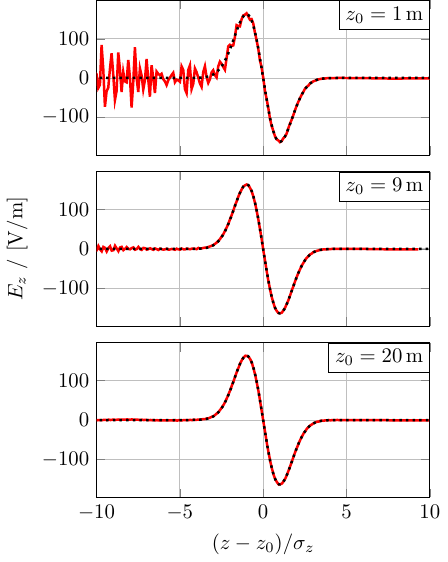} 
    \caption{Longitudinal electric field on-axis within the computational window at different bunch traveling distances $z_0$. The red curve is simulated; the black-dotted one is the analytical solution \eqref{eq_validation_EzTotal}.
    }
    \label{fig_validate_EzSeries}
\end{figure}

For the given bunch parameters, using a rectangular pipe with $w=\SI{100}{\mm}$, $h=\SI{15}{\mm}$ and a window length $l_\mathrm{w} = 20 \sigma_z$, we obtain a steady-state propagation distance of $l_\mathrm{t}\approx \SI{8.8}{\m}$. Figure \ref{fig_validate_EzSeries} shows $E_z$ on the beam axis at different bunch positions within the pipe. The presence of transient contributions in the simulated field (red curve) shortly after bunch injection is clearly visible. These contributions decay from the computational window as the bunch propagates and ultimately disappear at propagation distances larger than the estimated \SI{8.8}{\m}. The simulated steady-state field at \SI{20}{\m} agrees perfectly with the analytical solution \eqref{eq_validation_EzTotal} (black-dotted curve).

\begin{figure}[htb]
    \centering
    \includegraphics[width=0.85\linewidth]{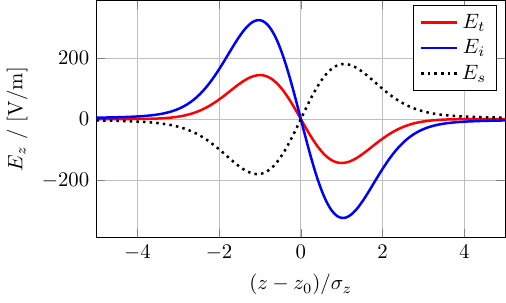}
    \caption{
        Contributions of the free-space SCF (blue) and scattered field (black-dotted) to the total longitudinal electric field along the axis (red) in a rectangular beam pipe.
    }
    \label{fig_validate_scattered}
\end{figure}

It is interesting to observe the relative contribution of the scattered field and that of the SCF to the total steady-state solution \eqref{eq_validation_EzTotal}. This is depicted in \figref{fig_validate_scattered}. The free-space SCF (blue curve) is computed according to \eqref{eq_validation_EzFreeSpace}. The total electromagnetic wakefield (red curve) is given by \eqref{eq_validation_EzTotal} and the scattered-field part (black dotted) by the difference between the two. As seen in \figref{fig_validate_scattered}, the scattered field tends to compensate the SCF of the particles. The resulting lower impedance of the beam in the pipe is often referred to in the literature as shielding effect \cite{Stupakov_2021}. The motivation for this terminology becomes obvious from the field separation approach \eqref{maxwell_ansatz} used in our formulation.

\subsection{Numerical accuracy}
\label{subsec:numConv}

To quantify the accuracy of simulations, we perform a numerical convergence analysis with respect to the mesh size used in the discretization of \eqref{eq_ScatField_HF}. The quantity of interest is the longitudinal {\em wake potential} per unit length \cite{Weiland_1992,Palumbo_1995}, which is given by the normalized electric field at steady state, $W (s) = E_{z} (s)/Q$. The relative rms error is then defined as
\begin{align}
    \varepsilon_{\mathrm{RMS}} = \sqrt{ \frac{\int \dd s \left( W (s)  - W_{\mathrm{ref}} (s)\right)^2}{\int \dd s W_{\mathrm{ref}}(s)^2}} , 
\end{align}
where $W_{\mathrm{ref}}(s)$ is the reference wake potential obtained from the analytical solution \eqref{eq_validation_EzTotal}.

\begin{figure}[htb]
    \centering
    \includegraphics[trim=0 0 0 1.5cm, clip, width=0.85\linewidth]{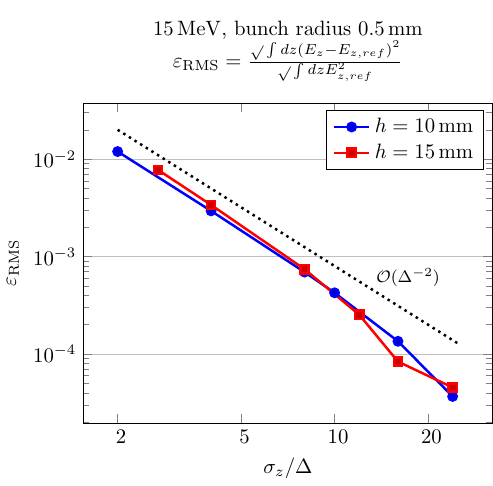}
    \caption{
        Relative rms error vs.\ mesh step size in the rectangular pipe case for two different pipe heights $h$.
    }
    \label{fig_validate_MeshConv_WakePot}
\end{figure}

\figref{fig_validate_MeshConv_WakePot} illustrates the numerical convergence in the case of a rectangular pipe of width $w=\SI{100}{mm}$ and two different heights $h$. The relative error is shown for different step sizes $\Delta$ of the mesh relative to the bunch length, where the mesh is uniformly refined in all three space directions. This error reduces to the per mille range at about 8 mesh cells per bunch length $\sigma_z$. For both considered pipes, the convergence rate is quadratic. The optimal convergence rate is due to the exact representation of the (piecewise) planar pipe geometry on the hexahedral mesh. In this case, the staircase approximation \eqref{magnetic_current1} and the boundary conformal approach \eqref{eq_ScatIpa_jmag} for the equivalent magnetic current on the pipe walls are equivalent.

\begin{figure}[htb]
    \centering
    \includegraphics[width=0.8\linewidth, trim=0 0 0 2.0cm, clip, ]{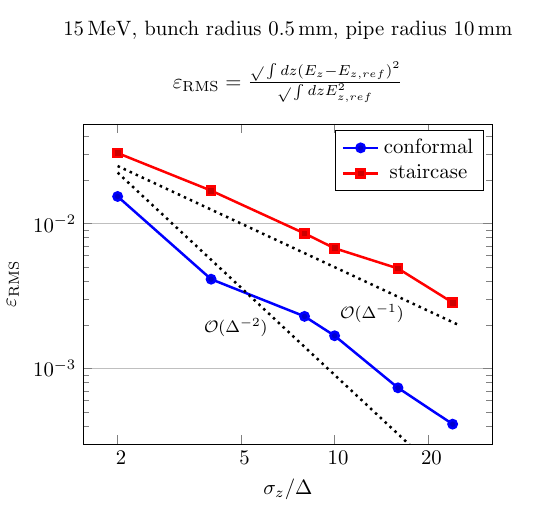}
    \caption{
        Relative rms error vs.\ mesh step size in the cylindrical pipe case using  staircase \eqref{magnetic_current1} and boundary conformal \eqref{eq_ScatIpa_jmag} discretization models, respectively.
    }
    \label{fig_validate_MeshConvCyl_WakePot}
\end{figure}

Next, we consider a circular pipe of radius $b=\SI{10}{\mm}$. The numerical convergence in this case is shown in \figref{fig_validate_MeshConvCyl_WakePot}. We use the staircase approximation \eqref{magnetic_current1} as well as the boundary conformal approximation \eqref{eq_ScatIpa_jmag} for the magnetic current. In the staircase model, the convergence rate is sub-quadratic. This is due to the poor representation of the curved pipe geometry on the hexahedral mesh. The second order convergence is fully recovered when the conformal boundary approach introduced in Section \ref{subsec:scat-iq} is used. For the same mesh resolution, this results in a substantially smaller numerical error compared to the case of the staircase discretization.

\section{RF photo-gun simulation}
\label{sec:gun_simulation}

In the following, we present a simulation study of a 7-cell, S-band RF photo-gun, which is presently installed at the SuperKEKB electron-positron collider facility in Tsukuba, Japan \cite{Akai_2018}. The gun is specifically designed to generate very high-current electron beams with single-bunch charges of up to \SI{5}{nC} and transverse emittances as low as \SI{10}{mm\cdot mrad} \cite{Natsui_2013,Natsui_2016}.
The gun supports a (quasi) traveling-wave RF accelerating field, which accelerates the particles up to about \SI{15}{MeV}. In addition, the RF-field provides for embedded transverse focusing of the beam, which makes external solenoid systems superfluous.

\begin{figure}[hbt]
    \centering
    \includegraphics[width=1\linewidth]{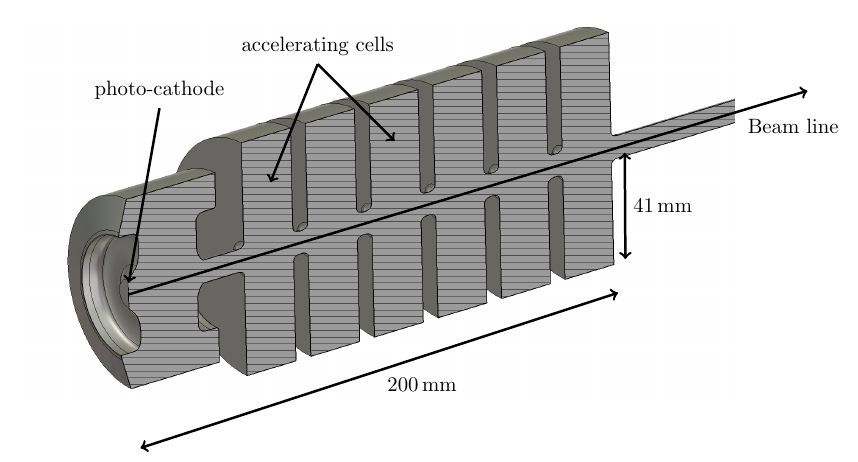}
    \caption{
    Cross-section of the geometry and main dimensions of the RF photo-gun at SuperKEKB (cf.\ \cite{Natsui_2013}).
    }
    \label{fig_gun_outline}
\end{figure}

Electromagnetic wakefields are usually not an issue in high-energy RF-guns due to their short length, typically consisting of 1-2 accelerating cells. However, the high beam current together with the multi-cell geometry of the SuperKEKB-design raises concerns regarding the impact of geometric wakefields on the beam dynamics, which is also the main motivation for this study.
The geometry of the gun is depicted in \figref{fig_gun_outline}. The main operation parameters used in our simulations are summarized in \tabref{tab_gun_modelParams}. Electron emission occurs from the photo-cathode, located at the backplane of the gun. The RF-field seen by the particles corresponds to the lowest longitudinal eigenmode of the cavity. We precompute it by means of a standard eigenmode solver \cite{CST}. Then, we incorporate this mode in the simulations as an externally applied field acting on the particles at the correct accelerating phase. 
\begin{table}
    \begin{tabularx}{1.0\linewidth}{l X}
        \toprule
        \textbf{Parameter} & \textbf{Value} \\
        \midrule
        \hline
        Peak accelerating field & \SI{72}{\mega\volt\per\meter} \\
        RF frequency & \SI{2.895}{\GHz} \\
       RF phase & \SI{34.9}{\degree} \\
        \hline
        Bunch charge & \SI{5}{\nano\coulomb} \\
        Initial distribution & radially uniform, $a=\SI{3.6}{\mm}$ \\
        Initial bunch length & \SI{20.0}{\ps} \\
        Final energy & \SI{14.2}{\MeV} \\
        \hline
    \end{tabularx}
    \caption{Beam and gun operation parameters used in the simulations.}
    \label{tab_gun_modelParams}
\end{table}

\subsection{Results}

\begin{figure}[hbt!]
    \centering
    \includegraphics[width=0.85\linewidth]{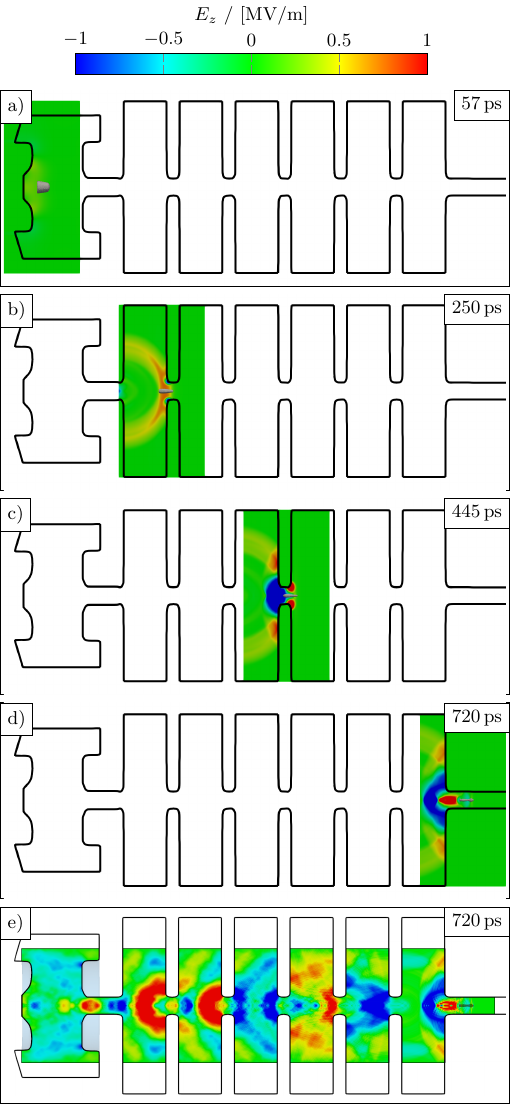}
    \caption{Electromagnetic wakefield in the gun at different times after emission. The evolving shape of the particle bunch is shown in gray. a)-d) Simulation using the scattered-field approach \eqref{eq_ScatField_HF} on a moving computational window (\wakeFMM). e) EM-PIC solution obtained with CST Particle Studio on a transversely truncated mesh (\wakeCST).}
    \label{fig_gun_snapshots}
\end{figure}

The basic result of an electron gun simulation is depicted in \figref{fig_gun_snapshots}, where the wakefield patterns excited by the particle beam within the gun at different time instances are shown. Figures \ref{fig_gun_snapshots}a)-d) show the $z$-component of the electric field computed with the scattered-field approach \eqref{eq_ScatField_HF} on a moving computational window. Figure \ref{fig_gun_snapshots}e) is the reference solution obtained by a full EM-PIC simulation using the commercial code CST Particle Studio \cite{CST}. Note that this latter simulation requires a full-size discretization of the gun geometry, which results in an extremely large computational mesh (see discussion at the end of this section). Therefore, we had to truncate the mesh in the two transverse directions to about 2/3rds of the cavity size (see\ \figref[e)]{fig_gun_snapshots}) in order to be able to run the simulations.

Visual inspection of the last two frames (\figref[d)]{fig_gun_snapshots}) and \figref[e)]{fig_gun_snapshots}) indicates that the field patterns at the end of the gun resulting from the two simulation approaches coincide closely. The same is true for the final shape of the particle bunch. 
However, in order to have a more detailed comparison including the impact of wakefields on the phase space of the bunch, we have performed a number of simulations under different physical assumptions, which we abbreviate in the following as:
\begin{itemize}[parsep=0em,leftmargin=*]
    \item
    \fsSC: Free-space simulation (no geometric wakefields) using the quasistatic approximation for the SCF \eqref{eq_Lorentz_transform}.
    \item 
    \fsCST: EM-PIC simulation in free-space (no geometric wakefields) using the CST solver.
    \item 
    \wakeFMM: Scattered-field formulation for the wakefields using the quasistatic approximation for the incident field computed with FMM.
    \item 
    \wakeLW: Scattered-field formulation for the wakefields using Liénard-Wiechert's solution \eqref{lw} for the incident field computation.
    \item 
    \wakeCST: Full EM-PIC simulation with the CST solver including (part of) the gun geometry.
\end{itemize}

\begin{figure}[hbt]
        \centering
        \includegraphics[width=0.8\linewidth]{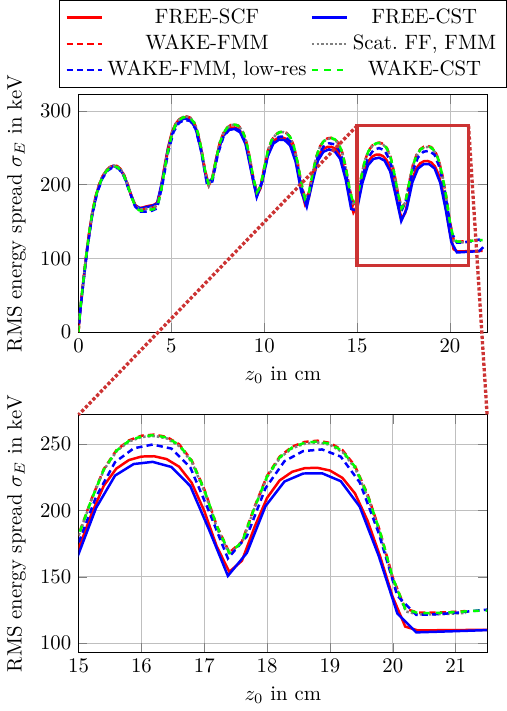}
    \caption{Energy spread of the particle bunch along the beam line computed with different simulation models. Bottom: enlarged view of the curves in the last \SI{6}{\cm} before the gun exit.}
    \label{fig_gun_RMSEnergySpread}
\end{figure}

\figref{fig_gun_RMSEnergySpread} shows the progression of the RMS energy-spread of the electron bunch within the gun cavity resulting from the different computational models. There is a very good agreement between the two free-space models, FREE-FMM and FREE-CST (solid lines), respectively. This confirms the validity of the quasistatic field approximation described in Section~\ref{subsec:green}, where geometric wakefield effects are neglected. Furthermore, it suggests that relativistic retardation and radiation effects (which are not included in FREE-FMM) for this particular application are expected to be weak.

Including now the gun geometry in the simulations, we observe that the energy spread of the bunch at the gun exit increases by about \SI{14}{\percent} compared to the free-space models, which is obviously due to geometric wakefields. This result is consistent for all three wakefield simulation models, WAKE-FMM, WAKE-LW and WAKE-CST. This agreement is remarkable given the complexity of the problem and the very different numerical models used in each of the simulations. 
 
In the \wakeCST simulations, we employed a computational mesh consisting of about \num{326e6} cells with a transverse grid size of \SI{0.14}{\mm} and a longitudinal grid size varying between \SI{0.075}{\mm} at the cathode and \SI{0.1}{\mm} after the first few \si{\cm}. The simulation required \SI{160}{\giga\byte} of RAM and a simulation time of \SI{\sim 10}{\hour} runtime on a multi-core machine. The \wakeFMM simulation converges already for a grid size of \SI{0.14}{\mm} in longitudinal and \SI{0.2}{\mm} in transverse directions, respectively. This results in a total of \num{8.8e6} mesh cells in the moving wakefield window, while the computational cost for the solution of the SCF problem is comparatively small. The overall \wakeFMM simulation requires \SI{\sim 8}{\giga\byte} of RAM and less than \SI{1.5}{\hour} of runtime on the same machine.

In general, it is difficult to quantify numerical efficiency based on simulation run times alone. This is because of the specific optimization details used in different code implementations. However, the above comparison demonstrates the pitfalls of classical EM-PIC simulations for this class of applications. The \wakeCST simulation requires the discretization of the full cavity structure. In addition, a finer mesh than in \wakeFMM is needed in the transverse directions, in order to resolve the charge distribution of the bunch. The same applies to the longitudinal direction in the region close to the cathode. This leads to a much denser mesh than it is actually needed for the wakefield problem. To maintain numerical stability, the time step of the simulation must be reduced correspondingly. Therefore, in EM-PIC simulations, the essentially quasistatic SCF of the bunch is updated much more frequently than necessary, thus, further increasing computational costs.
By contrast, the application of dedicated solvers in \wakeFMM allows to use different spatial and temporal resolutions, which are adapted to the scales of the SCF and wakefield problems, respectively. Thus, a higher numerical efficiency is achieved, while maintaining simulation accuracy.

\subsection{Discussion on relativistic effects}

\begin{figure}[hbt]
    \centering
    \includegraphics[width=0.8\linewidth, page=2]{wakes_RMS_devFMM-LW.pdf} 
    \includegraphics[width=0.8\linewidth, page=1]{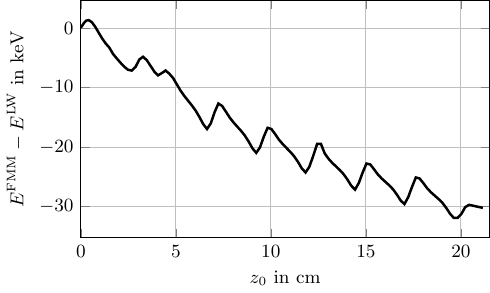} 
    \caption{
    Differences in energy spread, $\sigma_E$, (top) and mean energy, $E$, (bottom) between simulations with a quasistatic incident field (\wakeFMM) and a fully relativistic one (\wakeLW).
    }
    \label{fig_gun_relative_RMSEnergySpread}
\end{figure}

\begin{figure}[hbt]
    \raggedright
     \includegraphics[width=0.9\linewidth]{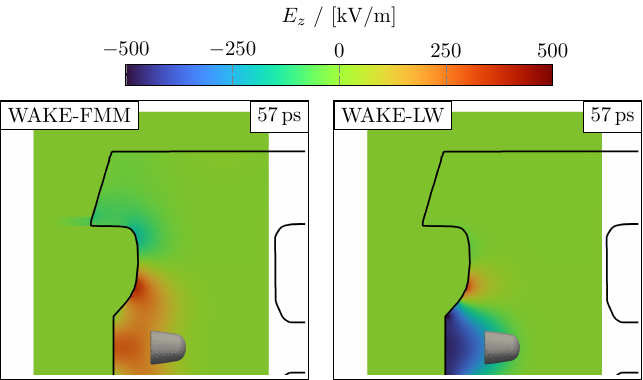} 
    \caption{
   Longitudinal wakefield pattern at \SI{60}{ps} after bunch emission. Left: using \wakeFMM. Right: using \wakeLW.
    }
    \label{fig_gun_fmm-lw}
\end{figure}

As seen in \figref{fig_gun_RMSEnergySpread}, relativistic radiation and retardation effects in the electron gun are negligible, as far as the energy spread of the beam is concerned. Nevertheless, we take a closer look at the simulation results in order to demonstrate the capability of the method to describe these effects. Figure \ref{fig_gun_relative_RMSEnergySpread} depicts the deviations of the RMS energy spread and the mean particle energy, when computed with a quasistatic incident field (\wakeFMM) and a fully relativistic one (\wakeLW). The largest difference in energy spread is observed within the first \SI{5}{cm} after the cathode, where the particles experience the strongest acceleration while undergoing a transition to the relativistic regime. Since \wakeFMM does not account for the retardation of the incident field, the wakefield at the boundary is excited instantaneously. Thus, this approach initially overestimates the energy spread induced on the beam. The wakefield computed with \wakeLW is of similar magnitude. However, it appears with a time delay due to retardation. This is illustrated in \figref{fig_gun_fmm-lw}, where the longitudinal field solutions shortly after bunch emission are shown. Obviously, the phase of the wakefield predicted in the two simulations is different. This explains the first peak in \figref{fig_gun_relative_RMSEnergySpread}(top). However, once the retarded wakefield excited at the cathode backplane catches up with the particles, the same amount of energy spread as in \wakeFMM is induced, thus, the difference between the two results disappears. This behavior is observed every time the bunch passes by an iris of the gun cavity. The wakefield effect in \wakeFMM is locally overestimated, however, the final energy spread at the gun exit predicted by both simulations is nearly the same.

A different picture is observed for the mean particle energy in \figref{fig_gun_relative_RMSEnergySpread}(bottom). The energy loss of the particles to the wakefield depends on the time of interaction. Since in the relativistic approach, the wakefields catch up later with the beam, the particles 'see' the wakefield at a higher energy. Thus, for the same momentum transfer, the change in particle energy is lower than in the quasistatic approach. The energy loss of the beam is systematically overestimated in \wakeFMM. However, the difference observed between the two simulations is in the few per mille range relative to the final beam energy of \SI{14.2}{MeV}. Thus, for all practical purposes, the quasistatic approximation of \wakeFMM is sufficient for this type of application.

\begin{figure}[hbt]
    \centering
    \includegraphics[scale=0.9]{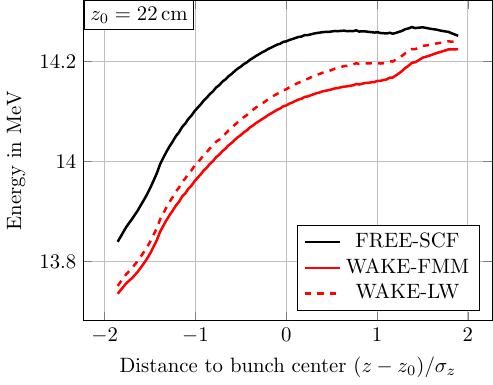}
    \caption{Longitudinal phase space of the bunch at the gun exit computed with different simulation approaches.
    }
    \label{fig_gun_sliceEnergy}
\end{figure}

A more detailed view of the longitudinal phase space of the beam at the gun exit is shown in \figref{fig_gun_sliceEnergy}. Three simulation models, \fsSC, \wakeFMM and \wakeLW, are compared.
The overall energy loss due to geometric wakefields in the gun is less than \SI{200}{keV} and, therefore, quite small compared to the total particle energy. This is even more so in the \wakeLW simulation, where the relativistic retardation of the wakefields is taken into account. Nevertheless, referring to \figref{fig_gun_RMSEnergySpread}, the wakefield-induced energy spread of \SI{\sim 14}{\percent} represents a sizable effect, which may impact beam quality in the downstream accelerator.

\section{Conclusion}

The main result of the paper is the scattered-field formulation of a numerical approach for self-consistent electromagnetic and beam dynamics simulations for particle accelerator applications. Except for the SCF particle interaction, this approach fully accounts for the geometric wakefields induced by the particle beam within the accelerator cavity. Formulating the beam-driven Maxwell's equations in terms of scattered fields, we are able to separate the temporal and spatial scales corresponding to space charge and wakefields, respectively, where the two parts of the problem are coupled by an equivalent magnetic boundary current.
For the computation of these currents, we introduce a boundary conformal procedure, which is shown to be optimally convergent with respect to mesh resolution. We compared numerical simulations with the analytical solution for a homogeneous beam pipe and found excellent agreement between the two.
The strength of this approach consists in the flexibility of choice that the formulation provides for the solutions of the subproblems, i.e., the SCF and wakefield, respectively. For each of the subproblems, specialized techniques can be used, which are optimized for accelerator applications. For example, for the computation of space charge boundary fields as well as for the inter-particle interaction, a quasistatic approximation based on the free-space Green's function method can be used. To account for radiation and retardation effects, a fully relativistic Liénard-Wiechert solution for the space charge problem can be employed.
We show that this reduces the computational cost of such simulations drastically compared, e.g., to conventional electromagnetic PIC. We demonstrate this in the simulation of a multi-cell electron photo-gun installed at the SuperKEKB accelerator facility.
For this particular application, we demonstrated an increase of about \SI{14}{\percent} in the energy spread of the electron bunch corresponding to a sizable deterioration of beam quality, which is solely due to electromagnetic wakefields. This underscores the importance of wakefield effects in electron gun applications, which are so far largely neglected in beam dynamics studies for particle accelerators. 

\section*{CRediT authorship contribution statement} 
J.C., E.G.\ and H.D.G.\ developed the main concept and wrote the manuscript. J.C.\ developed and implemented the scattered-field formulation and the boundary conformal approximation approach. J.C.\ performed all simulations presented in the paper. J.C.\ and E.G.\ designed and generated the figures. E.G.\ developed the wakefield solver, the Green's function, FMM and Liénard-Wiechert space charge solvers. E.G.\ acquired the funding for this project. J.C., E.G.\ and H.D.G.\ conducted the analysis of simulation results. All authors reviewed the manuscript.

\section*{Acknowledgements} 
This work has been funded by the Deutsche Forschungsgemeinschaft (DFG, German Research Foundation) – Project-ID 264883531 – GRK 2128 "AccelencE".





\end{document}